\documentclass[graybox]{svmult}

\usepackage{mathptmx}       
\usepackage{helvet}         
\usepackage{courier}        
\usepackage{type1cm}        
\usepackage{makeidx}         
\usepackage{graphicx}        
\usepackage{multicol}        
\usepackage[bottom]{footmisc}

\makeindex             

\begin{document}

\title*{Properties of holographic dark energy at the Hubble length}
\author{Ivan Duran and Luca Parisi}
\institute{Ivan Duran \at Universitat Aut\`{o}noma de Barcelona, \email{ivan.duran@uab.cat}
\and Luca Parisi \at Universit\`{a} di Salerno, \email{parisi@sa.infn.it}}
%
%
\maketitle

\abstract*{}

\abstract{We consider holographic cosmological models of dark
energy in which the infrared cutoff is set by the Hubble's radius.
We show that any interacting dark energy model, regardless of its
detailed form, can be recast as a non interacting model in which
the holographic parameter $c^{2}$ evolves slowly with time. Two
specific cases are analyzed. We constrain the parameters of both
models with observational data, and show that they can be told
apart at the perturbative level.}

\section{Introduction}\label{sec:Intro}

Whatever the nature of DE it seems reasonable that it fulfills the
holographic principle \cite{gerard-leonard}. Based on this, Li
\cite{MLi} proposed for the  density of DE the expression
\begin{equation} \rho_{X}= \frac{3 M_{P}^{2}\,  c^{2}}{L^{2}}\, . \label{eq:rhox}
\end{equation}
\noindent where $c^{2}$ is a dimensionless parameter and $L$ the
IR cutoff.

We will take $L$ as  the Hubble radius, $L=H^{-1}$, see e.g.
\cite{hubbleradius}.  See
e.g.\cite{MLi,gao,lxu,suwa,DuranPavonPrd} for other choices. It
has been argued that an IR cutoff defined by $H^{-1}$ cannot lead
to an accelerated Universe. However, if DM and DE interact
according to
\begin{equation}\label{eq:evolEqIntMX}
\dot{\rho}_{M} + 3H \rho_{M} = Q    \qquad {\rm and} \qquad
\dot{\rho}_{X} + 3H (1+w)\rho_{X} = - Q  \, ,
\end{equation}
\noindent where $Q>0$ is the interaction term, an accelerated
expansion can be achieved \cite{ZimdahlPavon}.

In \cite{Pavon:2005yx} the $c^2$ parameter was considered to increase slowly with time in such a
way that $0 < \left(c^{2}\right)\dot{}\leq H$. In what follows,
quantities referring to models with variable $c^{2}$ will be noted
by a tilde. By assumption their energy densities conserve
separately,
\begin{equation}\label{eq:EvolutionMX}
\dot{\tilde{\rho}}_{M}=-3H\tilde{\rho}_{M} \qquad {\rm and} \qquad
\dot{\tilde{\rho}}_{X}=-3H(1+\tilde{w})\tilde{\rho}_{X} \,.
\end{equation}
By considering both points of view it was demonstrated that
identical backgrounds evolutions can be described by an
interacting holographic DE model, with $c^{2}$ strictly fixed, or
by a non-interacting holographic DE model in which $\tilde{c}^{2}$
depends weakly on time \cite{DuranParisi}. In spite that the
global evolution  is identical in both scenarios, the energy
densities and the EoS parameters can behave rather differently.
\section{Proposed models: model 1 and model 2}\label{sec:models}
\begin{figure*}[!htb]
  \begin{center}
    \begin{tabular}{cc}
      \resizebox{50mm}{!}{\includegraphics{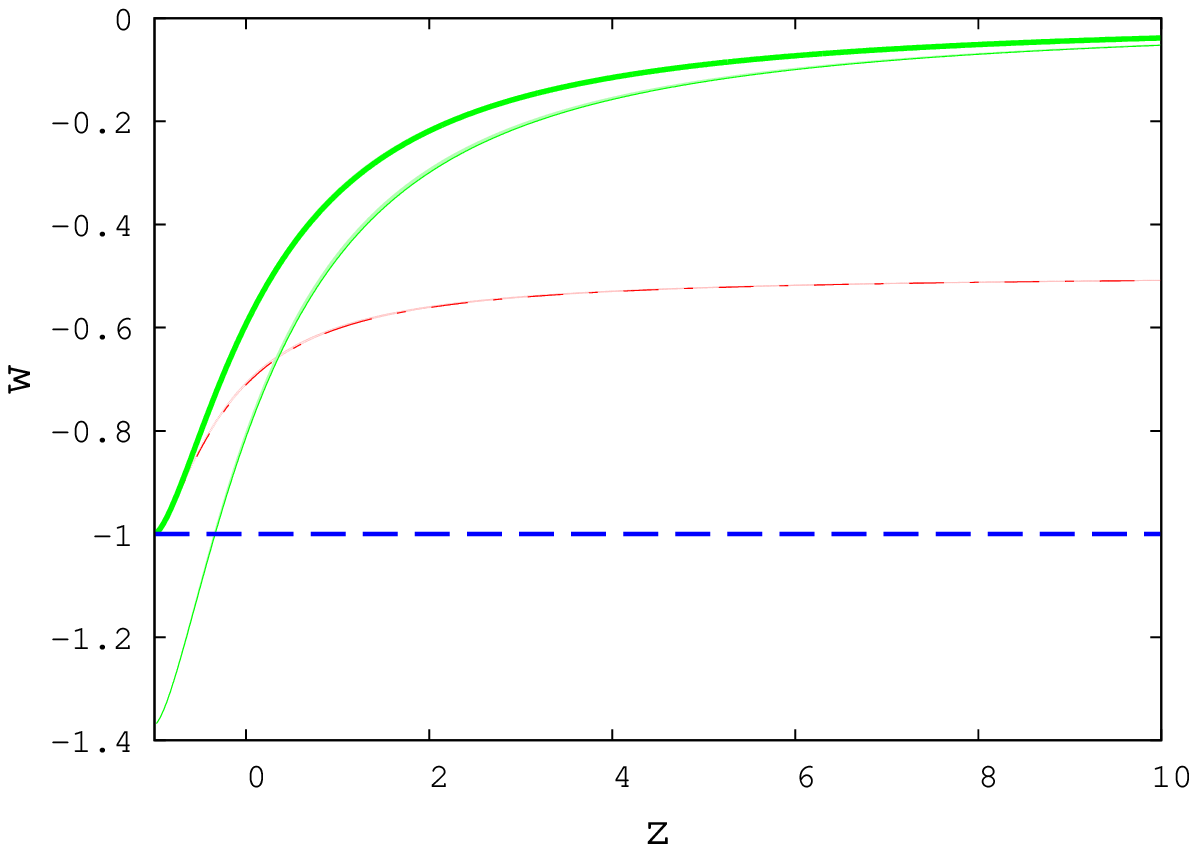}}&
      \resizebox{50mm}{!}{\includegraphics{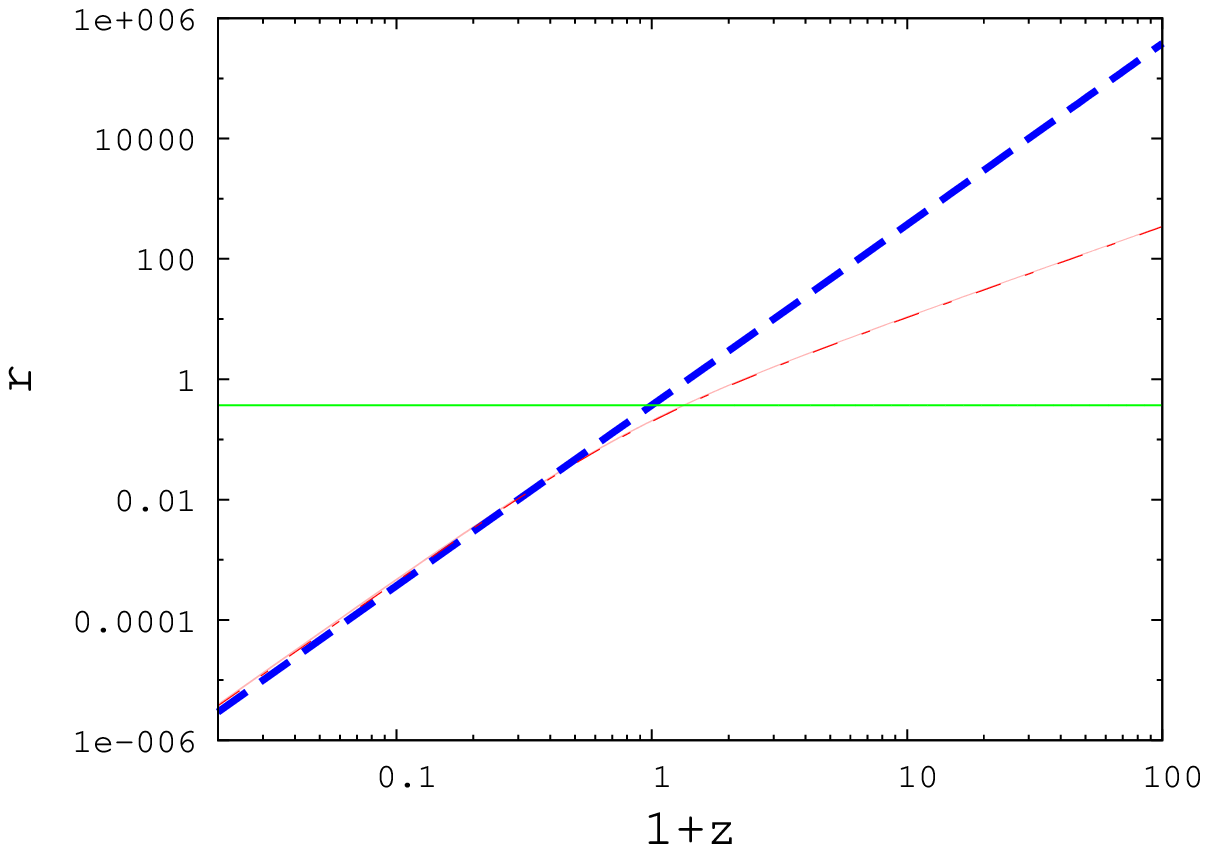}}\\
      \resizebox{50mm}{!}{\includegraphics{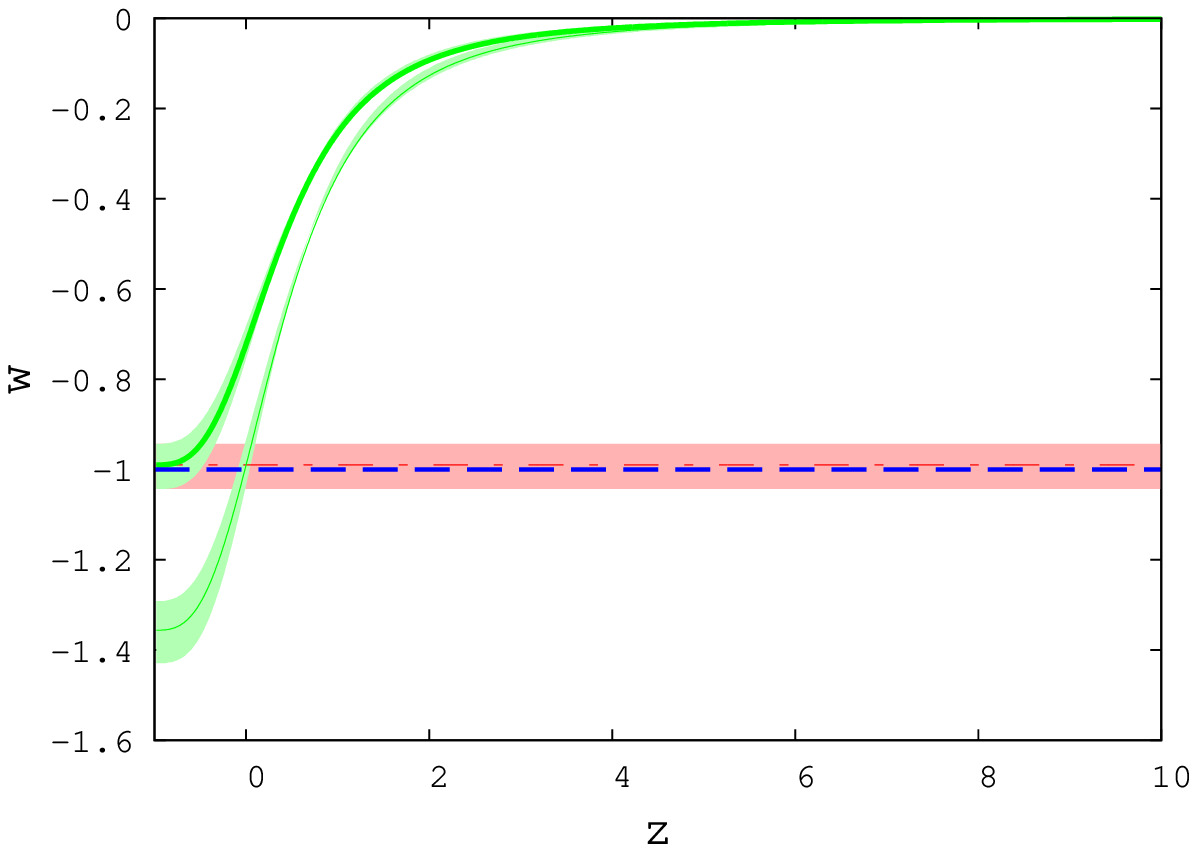}}&
      \resizebox{50mm}{!}{\includegraphics{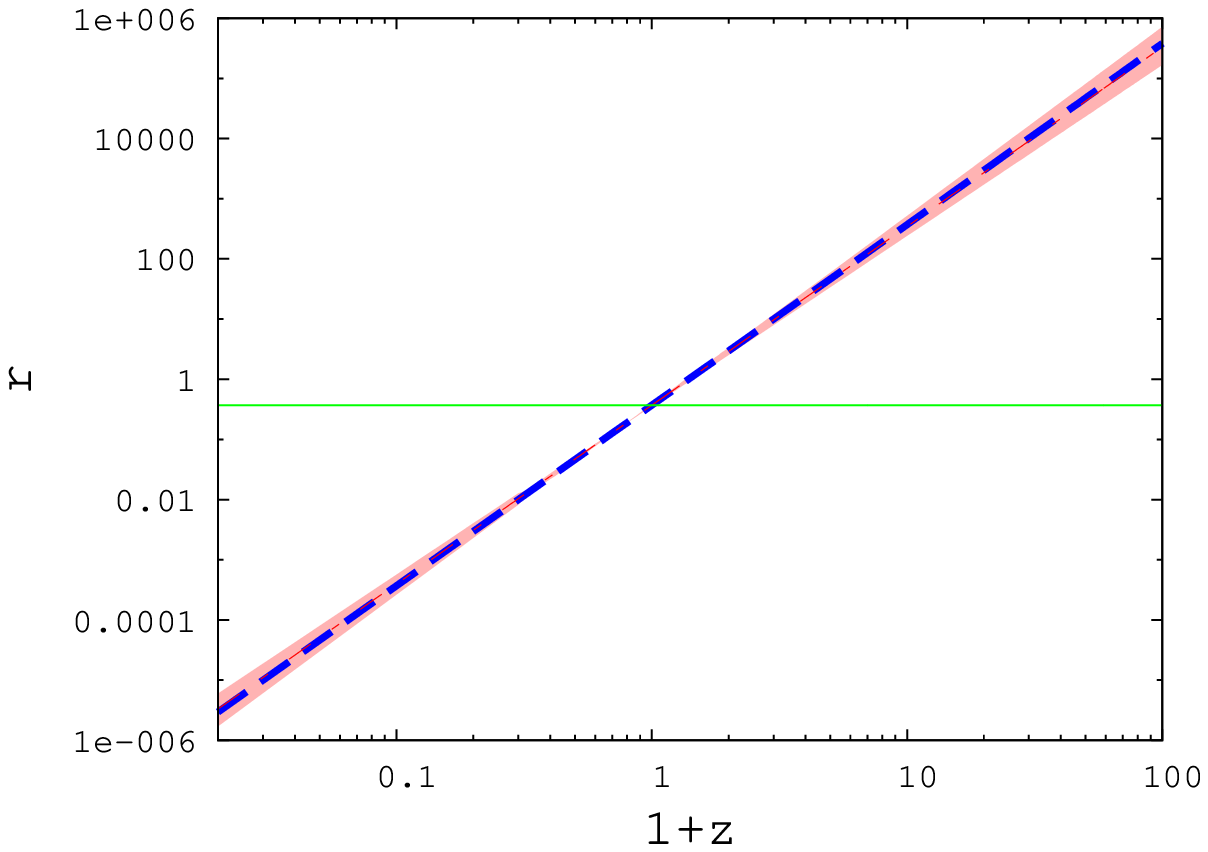}}\\
    \end{tabular}
    \caption{\scriptsize Top panels are for model 1 and bottom panels for model 2. Left panels: EoS parameter for
    the the interacting ($w$ thin line, and $w_{eff}$ thick line), the $\tilde{c}^{2}$  and $\Lambda$CDM models.
    Right panels: energy densities ratios, $r\equiv\rho_{M}/\rho_{X}$, versus $1+z$ for the $\Lambda$CDM,
    the interacting and the $\tilde{c}^{2}$ models. Solid (green) lines are used for the interacting case,
    thin dot dashed (red) lines for the $\tilde{c}^{2}$ model, and thick short dashed (blue) for $\Lambda$CDM.}
    \label{fig:ex1}
 \end{center}
\end{figure*}

Here we consider the holographic interacting model studied in
\cite {PavonDuranZimdahl} in order to construct its equivalent
$\tilde{c}^{2}(t)$ model. In the former the IR cutoff was also set
by the Hubble's length and the interaction term was $Q \equiv
3AH_{0}\rho_{M}$, with $A$ a semipositive definite constant,
related to the constant decay rate of DE into DM , $\Gamma$, by
$A\equiv\frac{\Gamma}{3H_{0}r}$, with $r \equiv
\rho_{M}/\rho_{X}$. Thus, the Hubble function takes the form
\begin{equation}\label{eq:H1}
H=H_{0}\left(A+(1-A)(1+z)^{\frac{3}{2}}\right)\, ,
\end{equation}
We expand $H^{2}(z)$ assuming that the  $(1+z)^{3}$ term
corresponds to DM and identify the remainder of the expression as
the DE energy density. Thus,
\begin{equation}\label{eq:rhoMNI1}
\frac{M_{P}^{-2}}{3H_{0}^{2}}\tilde{\rho}_{M} =(1-A)^{2}(1+z)^{3}
\qquad and \qquad \frac{M_{P}^{-2}}{3H_{0}^{2}}\tilde{\rho}_{X} = A^{2}+2\,A(1-A)(1+z)^{\frac{3}{2}}\, ,
\end{equation}
\noindent alongside with
\begin{equation}\label{eq:c2(z)}
\tilde{c}^{2}=\frac{2A(1-A)(1+z)^{\frac{3}{2}}+A^{2}}{\left(A+(1-A)(1+z)^{\frac{3}{2}}\right)^{2}}\, .
\end{equation}
The best fit values are found to be $H_{0}=69.4\pm 1.7$ and
$A=0.588 \pm 0.004$, while $\chi^{2}/dof=1.00$. For details see
\cite{DuranParisi}. As  the top right panel of Fig. \ref{fig:ex1}
shows, the coincidence problem (i.e., ``why the densities of DM
and DE are of the same order precisely today?") gets solved ($r$
stays constant) in the interacting case (solid green line). In the
$\tilde{c}^{2}$ model (thin dot-dashed red lines) it is not solved
but results much less severe than in $\Lambda$CDM (thick short
dashed blue line).
\\
We next propose  model 2. In this model DM and DE evolve
separately but $\tilde{c}^{2}$ varies slowly with time. In order
to have $0 \leq \tilde{c}^{2} \leq 1$, and
$\left(\tilde{c}^{2}\right)\dot{}\geq 0$ we define
\begin{equation}\label{eq:cDeT4}
\tilde{c}^{2}=\frac{1}{1+\tilde{r}_{0}(1+z)^{\epsilon}}
\end{equation}
\noindent where $\tilde{r}_{0}\equiv\frac{\tilde{\Omega}_{M0}}{\tilde{\Omega}_{X0}}$ and
$\epsilon$ a semipositive definite constant. In this case
\begin{equation}\label{eq:H4}
H=H_{0}\sqrt{\tilde{\Omega}_{M0}(1+z)^{3}+\tilde{\Omega}_{X0}(1+z)^{3-\epsilon}}
\end{equation}
\noindent is identical to a spatially flat $w$CDM model with
$\tilde{w}=-\frac{\epsilon}{3}$. If we consider Eq.(\ref{eq:H4})
as resulting from some interaction between DE and DM, the
interacting term would be
\begin{equation}\label{eq:Q4}
Q=-3\,c^{2}\,w\rho_{M}H \;,
\end{equation}
\noindent Detailed calculations can be found in
\cite{DuranParisi}. The best fit values are $\Omega_{X\,0}=0.73\pm
0.007$, $H_{0}=71.5 \pm 2.6$ and $ \epsilon = 2.97^{+ 0.16}_{-
0.14}$, being $\chi^{2}/dof=0.97$. As the bottom right panel of
Fig. \ref{fig:ex1} shows the interacting model (solid green line)
solves the coincidence problem.
\section {Evolution of the subhorizon perturbations}\label{Perturbations}
\begin{figure*}[!htb]
  \begin{center}
    \begin{tabular}{cc}
      \resizebox{50mm}{!}{\includegraphics{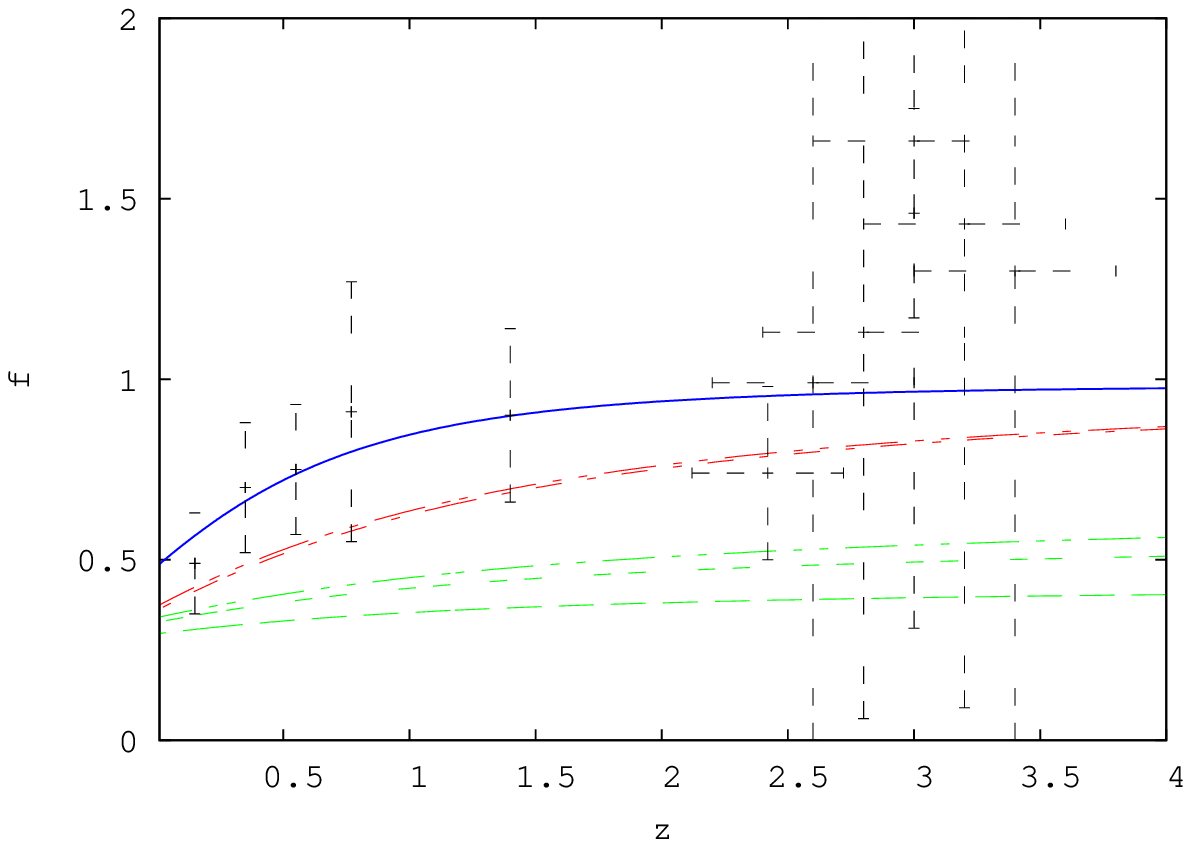}}&
      \resizebox{50mm}{!}{\includegraphics{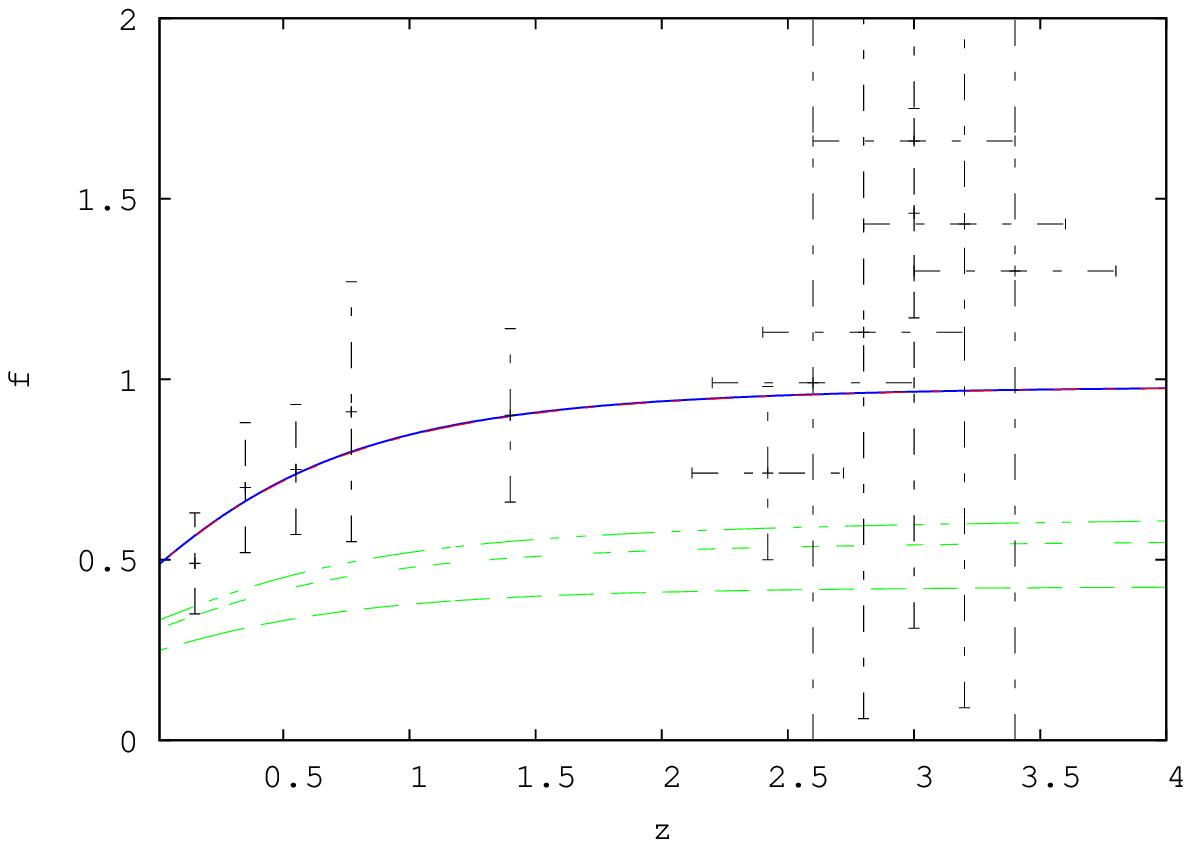}}\\
    \end{tabular}
    \caption{\scriptsize Left panel: evolution of the growth function, $f$, versus redshift
    for model 1. Right panel: the same for model 2. The dashed (green)
    lines describe the interacting scenario, the dot-dashed (red) lines the $\tilde{c}^{2}$,
    and the solid (blue) line the $\Lambda$CDM. The observational data were borrowed from \cite{Gong}.}
    \label{fig:f}
 \end{center}
\end{figure*}

In the interacting case, the energy-momentum tensors of DM and DE
are not independently conserved, $T^{\mu \,\nu}_{i\;\;
;\mu}=Q^{\nu}_{i}$. For subhorizon scales, i.e., $k \gg aH$, the
density and energy and momentum conservation equations simplify to
\begin{eqnarray}
\label{eq:PertrubedDensityNew}
\dot{\delta}_{M}&=&-\frac{\theta_{M}}{a}\qquad and \qquad \dot{\theta}_{M}=-H\theta_{M}+\frac{k^{2}}{a}\phi \\
\label{eq:PertrubedDEDensityNew}
\dot{\delta}_{X}&=&-\left(1+w\right)\frac{\theta_{X}}{a}-3H\left(1-w\right)\delta_{X}+\frac{1}{\rho_{X}}\left(Q\delta_{X}-\delta Q\right) \;,\\
\label{eq:PertrubedDEVelocityNew}    \dot{\theta}_{X}&=&\frac{1}{\left(1+w\right)}\frac{k^{2}}{a}\delta_{X}-\frac{Q}{\left(1+w\right)\rho_{X}}\left(\theta_{M}-2\theta_{X}\right)\;,
\end{eqnarray}
See \cite{DuranParisi} for the description of the $\delta Q$ in
each model. To confront it with observations, we resort to the
growth function, $f\equiv d \ln \delta_{M}/d \ln a$
\cite{Steinhardt}. We can see in Fig. \ref{fig:f}, that matter
density perturbations clearly different in the interacting and the
$\tilde{c}^{2}$ scenarios.
\begin{acknowledgement}We are indebted to Diego Pav\'{o}n , Gaetano Vilasi, Ninfa Radicella and
Fernando Atrio-Barandela for fruitful discussions. ID was supported by the Spanish MICINN under
Grant No. FIS2009-13370-C02-01, by the Generalitat de Catalunya under Grant No. 2009SGR-00164. LP
was partially supported by the Italian MIUR through the PRIN 2008 grant.
\end{acknowledgement}
%
%
%

\end{document}